\documentclass[12pt]{article}
\usepackage{graphicx}

\newcommand\pubnumber{DPF2015-17}
\newcommand\pubdate{November 1, 2015}

\def\caiaddress{Department of Physics\\
Queen's Univesity, Kingston, Ontario K7L 3N6, Canada}
\def\caiemail{\footnote{Email: beicai@owl.phy.queensu.ca}}

\def\Title#1{\begin{center} {\Large #1 } \end{center}}
\def\Author#1{\begin{center}{ \sc #1} \end{center}}
\def\Address#1{\begin{center}{ \it #1} \end{center}}

\newcommand\pubblock{\rightline{\begin{tabular}{l} \pubnumber\\
         \pubdate  \end{tabular}}}
\newenvironment{Abstract}{\begin{quotation}  }{\end{quotation}}
\newenvironment{Presented}{\begin{quotation} \begin{center}
             PRESENTED AT\end{center}\bigskip
      \begin{center}\begin{large}}{\end{large}\end{center} \end{quotation}}
\def\Acknowledgments{\bigskip  \bigskip \begin{center} \begin{large}
             \bf ACKNOWLEDGMENTS \end{large}\end{center}}

\def\beq{\begin{equation}}
\def\eeq#1{\label{#1}\end{equation}}
\def\eeqn{\end{equation}}

\def\beqa{\begin{eqnarray}}
\def\eeqa#1{\label{#1}\end{eqnarray}}
\def\eeqan{\end{eqnarray}}

\let\bar=\overbar

\def\Dslash{\not{\hbox{\kern-4pt $D$}}}
\def\dslash{\not{\hbox{\kern-2pt $\del$}}}

\def\msb{{\bar{\ssstyle M \kern -1pt S}}}

\begin{document}
\begin{titlepage}
\pubblock

\vfill
\Title{The DEAP-3600 Dark Matter Experiment}
\vfill
\Author{ Bei Cai\caiemail~for the DEAP Collaboration}
\Address{\caiaddress}
\vfill
\begin{Abstract}
The DEAP-3600 experiment uses 3.6~tons of liquid argon for a sensitive dark matter search,
with a sensitivity to the spin-independent WIMP-nucleon cross-section of $10^{-46}$~cm$^2$
at 100~GeV WIMP mass. This high sensitivity is achievable due to the large target mass
and the very low backgrounds in the spherical acrylic detector design as well as at the
unique SNOLAB facility in Sudbury, Canada. Pulse shape discrimination is used
to reject electromagnetic backgrounds from the WIMP induced nuclear recoil signal. We
started taking commissioning data in early 2015 with vacuum and later gas inside the
detector. Argon fill is expected in winter 2015. An overview and status of the DEAP-3600 experiment are presented in this paper, with an emphasis on control and mitigation of detector backgrounds.
\end{Abstract}
\vfill
\begin{Presented}
DPF 2015\\
The Meeting of the American Physical Society\\
Division of Particles and Fields\\
Ann Arbor, Michigan, August 4--8, 2015\\
\end{Presented}
\vfill
\end{titlepage}
\def\thefootnote{\fnsymbol{footnote}}
\setcounter{footnote}{0}

\section{Introduction}
It is now well known that dark matter makes up about one quarter of our universe, and accounts for more than five times the energy density of normal matter. The origin of dark matter is currently one of the most important questions in particle astrophysics. One favoured hypothesis is that this dark matter is comprised of Weakly Interacting Massive Particles or WIMPs. The DEAP collaboration has designed and built the DEAP-3600 detector at SNOLAB, 2~km underground in the Creighton Mine in Sudbury, Canada. DEAP-3600 is one of the most sensitive experiments for direct detection of dark matter with a sensitivity to the spin-independent WIMP-nucleon cross-section of $10^{-46}$~cm$^2$ at 100~GeV WIMP mass.

\section{The DEAP-3600 detector}
\begin{figure}[htb]
\centering
\includegraphics[width=2.8in]{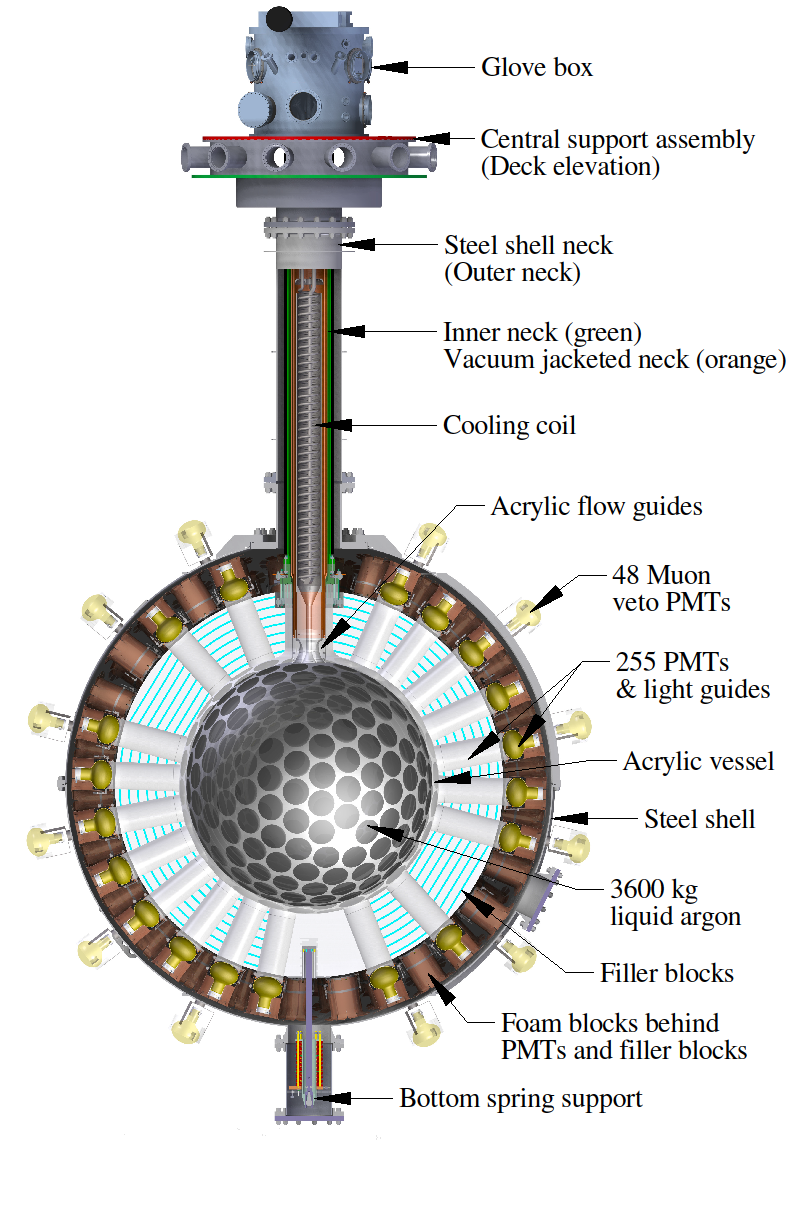}\vspace*{-0.25in}
\caption{Central parts of the DEAP-3600 detector.}
\label{fig:detector}
\end{figure}

The central parts of the DEAP-3600 detector are shown in Figure~\ref{fig:detector}. 3.6~tons of liquid argon target is enclosed in a 2-inch thick ultrapure acrylic vessel (AV) with inner radius of 85~cm. The inside is coated with tetraphenyl
butadiene (TPB, $\rm C_{28}H_{22}$) wavelength shifter, which serves
to shift the ultraviolet (UV) light generated by argon scintillation
to the visible region. 255 high quantum efficiency photomultiplier tubes (Hamamatsu R5912) collecting scintillation light are separated from the argon through acrylic light guides and polyethylene filler blocks, providing neutron shielding and thermal insulation. A large stainless steel shell encloses all the inner detector components, which itself is instrumented with veto PMTs and immersed in an 8-meter diameter ultrapure water tank for radiation shielding as well as \v{C}erenkov veto for cosmogenic muons. The detector is cooled through liquid nitrogen filled cooling coil installed in the neck with acrylic flow guides attached to the bottom to guide warm argon flow. The glovebox on top of the neck allows insertion or extraction of the process systems, resurfacer, TPB deployment system and calibration sources in a radon-free environment.

The AV acrylic was produced from pure MMA monomer at Asia plant of Reynolds Polymer Technologies (RPT) with strict control of radon exposure. It was then cast into panels that were thermoformed to orange slices, which were then bonded to form the sphere. The acrylic vessel was built in three pieces: AV main sphere, collar and neck, and they were bonded together underground at SNOLAB. A rotating support was used to allow manipulation of the AV for light guide bonding and machining. The AV went through multiple annealing processes in which the temperature of the acrylic was heated to 80-85$^{\circ}$C in a custom built annealing oven to reduce stress in the acrylic induced by bonding and machining. After the construction of the AV was completed, the PMTs were mounted onto the light guides and the gaps in between the light guides were filled with polyethylene filler blocks that were pre-assembled in a clean room. A stainless steel shell encloses and supports the weight of the completed inner detector that includes liquid argon, AV, light guides, filler blocks, PMTs and detector cabling. The steel shell also serves as a water-tight, light-tight and safety container.

\section{Backgrounds mitigation}
The detector design takes advantage of extensive ultra-low background techniques to allow for less than one event from background sources in 1,000,000 kg-days of exposure. Backgrounds from beta decay of $^{39}$Ar can be reduced using pulse shape discrimination (PSD) since the time evolution of the scintillation light pulse induced by an electromagnetic particle in liquid argon is very different from nuclear recoils~\cite{Boulay2006}. Low background gamma assay and radon emanation measurements were used in selection of only low-radioactivity materials for all the DEAP-3600 detector components, which allowed for great reduction of neutron backgrounds from uranium fission and ($\alpha$, n) reactions.

$^{210}$Pb level in the AV acrylic was measured by vaporizing 2~kg of acrylic in a dedicated acrylic vaporization system, collecting $^{210}$Pb from the residue and counting in a germanium well detector and an alpha counter separately. The measurements are consistent with all relevant backgrounds, giving an upper limit of 2.2$\times10^{-19}$~g/g $^{210}$Pb~\cite{Nantais}.

\begin{figure}[htb]
\centering
\includegraphics[width=4.5in]{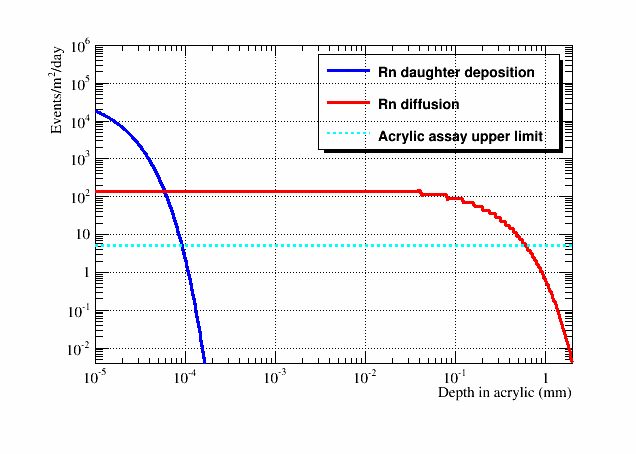}\vspace*{-0.25in}
\caption{Expected alpha event rate in the $^{210}$Pb decay chain in DEAP-3600 acrylic vessel before resurfacing. Contributions from radon daughter deposition and radon diffusion are shown in blue and red respectively. The cyan line is from measured $^{210}$Pb upper limit. The resurfacer took away $\sim$0.4 mm of acrylic and the residue activity on the AV inner surface is $\sim$10~$\alpha$/m$^2$/day.}
\label{fig:210Pb}
\end{figure}

A process flange sealing the top of the AV was installed right after the AV neck was bonded, which allowed purging the AV volume with radon-free gas during and after the annealing processes. We recorded the radon exposure history of the AV which allowed for calculation of the acrylic radiopurity before resurfacing. In total the AV was exposed to radon-laden air for a total of 9~months on surface, 6~months in underground mine air and 1~month in radon reduced air at radon rates of $\sim$12, 130 and 10~Bq/m$^3$, respectively. Figure~\ref{fig:210Pb} shows the $^{210}$Pb alpha activity in the AV as a function of depth before the resurfacing took place. This gives $\sim$5$\times$$10^4$~$\alpha$/m$^2$/day on AV inner surface from radon diffusion as well as radon daughter deposition, which quickly drops down since radon daughter nuclei do not travel far into acrylic.

\begin{figure}[htb]
\centering\hspace*{+0.6in}
\includegraphics[width=2.8in]{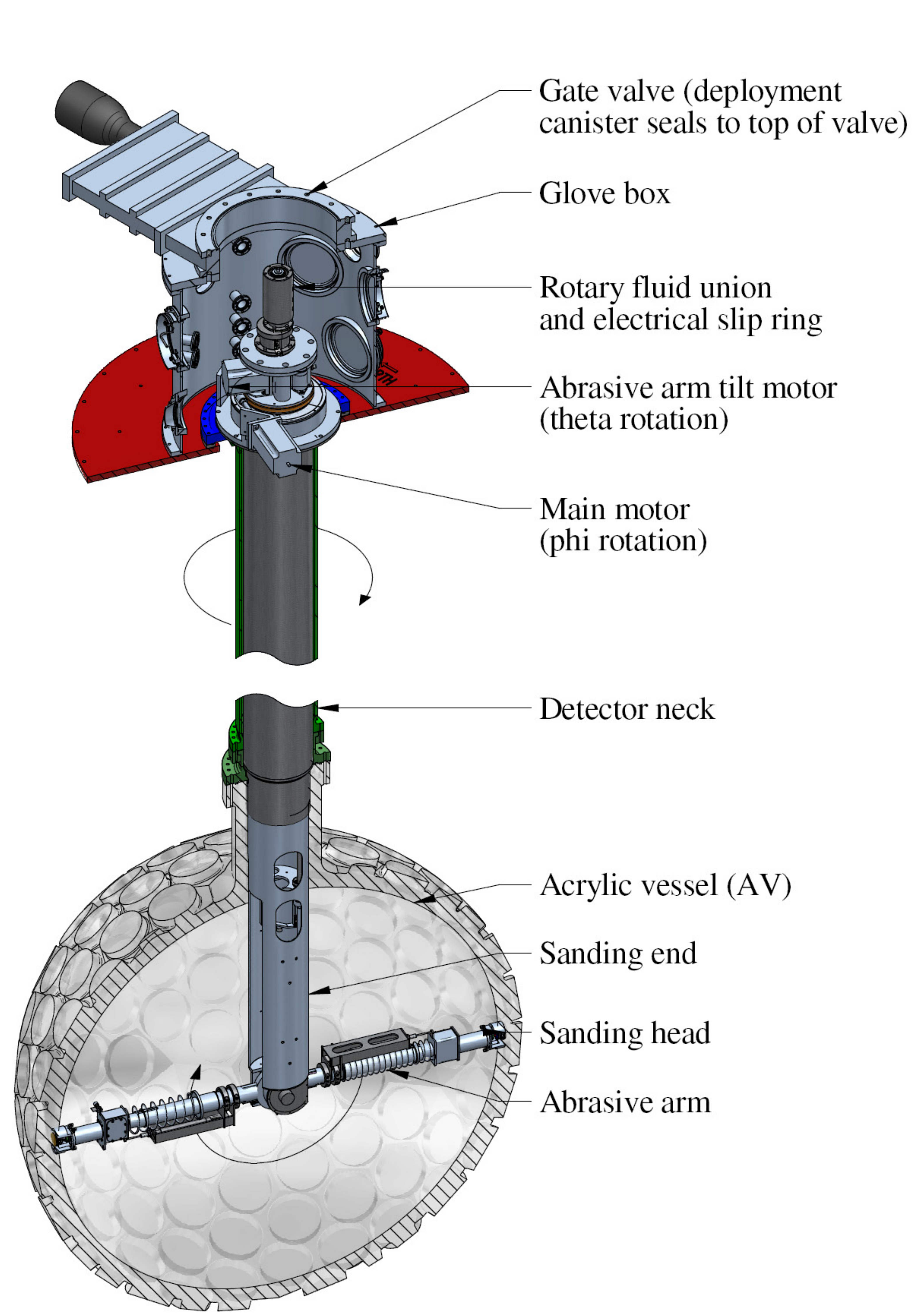}\vspace*{+0.1in}
\caption{Overview of the resurfacer device when deployed in the AV.} \label{fig:resurfacer}
\end{figure}

A resurfacing device (shown in Figure~\ref{fig:resurfacer}) was designed to remove the inner radioactive surface layer of the spherical AV. The resurfacer motor assembly and sanding end are connected via a body extension tube and vertical drive shaft extension, forming an assembly that is 20-foot long. The materials used in the fabrication of the resurfacer are of low radioactivity. Stainless steel is used for all mechanical housings, due to its low radiopurity and resistance to corrosion. Prior to the full resurfacer assembly underground at SNOLAB, the resurfacer was commissioned at Queen's University using a shorter prototype device that used the final drive end and sanding end of the resurfacer, however excluded the body tube extension and drive shaft extension. Sanding tests were performed on two acrylic panels installed on the north and south poles. The acrylic removal rates and durability of the sanding pad (3M Flexible Diamond QRS Cloth Sheet 6002J M74) were measured. The resurfacer was then moved to SNOLAB, reassembled with the extension pieces and tested before deployment inside the AV.

The resurfacer was deployed into the AV in September 2014. It was operated with the detector volume hermetically sealed from the lab and continuously purged with purified nitrogen gas. Ultrapure water was constantly sprayed onto the AV and sediment created during sanding was continuously extracted through a suction line adjacent to the sanding head. The AV was resurfaced for a total of 200~hrs, equivalent of 27 passes all across the inner surface. All radon daughters deposited on the AV surface were removed and the estimated alpha event rate in the $^{210}$Pb decay chain is $\sim$10~$\alpha$/m$^2$/day on the AV surface after resurfacing. With the sanding heads retracted, the AV was flushed with ultrapure water for multiple passes. The resurfacer was extracted through the glovebox and into an extraction canister. The AV, glovebox and extraction canister maintained nitrogen purge flow and were hermetically sealed from the lab during resurfacer extraction. The AV volume has since been either kept in vacuum or purged with radon-scrubbed nitrogen gas purge with no exposure to radon.

\section{Current status and summary}
The TPB wavelength shifter deposition source was deployed into the center of the acrylic vessel and heated to high temperature to allow water vapor in the acrylic to outgas. TPB was then loaded into the source, heated and evaporated onto the inner surface of acrylic vessel. A diffuse laser ball was deployed inside the AV and data with lasers at 3 different wavelengths (375~nm, 405~nm and 445~nm) were taken with the laser ball at various $z$ positions to calibrate the PMT efficiency and TPB coating uniformity. Timing calibration and detector operation stability have been monitored weekly by an acrylic reflector and fiber optics (AARF) system, where flashes of light from 435-nm LEDs driven by fast electronics are guided into the detector by acrylic fibers attached onto 20 lights guides uniformly spread across the detector and 2 fibers attached onto the acrylic neck area.

The water shield tank was gradually filled with ultrapure water to the top and we observed reduction of detector event rate (dominated by Cerenkov light) as the water level in the shield tank increased. This is what we expected since the neutrons and gamma rays coming from the decays of uranium and thorium in the rock are absorbed in the water shield. The cooling coil and acrylic flow guides have now been installed in the detector neck and leak checking is ongoing. We anticipate start of argon gas and liquid fill before the end of this year.

In summary, significant efforts have been put in to minimize DEAP-3600 detector backgrounds, including quality control of ultra-pure acrylic vessel manufacture and resurfacing before argon fill, selection of low radioactivity materials and limiting exposure to radon for detector components during assembly and construction. The PMT system has been operational since the end of 2014 and is being calibrated with light sources. We plan to commission the detector with argon gas followed by cool down/liquid argon fill before the end of 2015. DEAP-3600 detector is expected to have a leading sensitivity after 2 months of liquid argon data taking. After three tonne-year exposure we expect to reach our peak sensitivity to the WIMP-nucleon cross-section of $10^{-46}\ \rm cm^{2}$ at 100~GeV WIMP mass.

\newpage

\Acknowledgments
This work is supported by the National Science and
Engineering Research Council of Canada (NSERC), by
the Canada Foundation for Innovation (CFI), by the Ontario
Ministry of Research and Innovation (MRI) and by
the European Research Council (ERC). We thank Compute
Canada, Calcul Qu\'ebec, McGill University's centre for High Performance Computing and the High Performance
Computing Virtual Laboratory (HPCVL), for computational support and data storage.
We are grateful to SNOLAB and Vale Canada, Ltd. for excellent on-site
support.


\begin{thebibliography}{99}


\bibitem{Boulay2006} M.G. Boulay and A. Hime, ``Technique for direct detection of weakly interacting massive particles using scintillation time discrimination in liquid argon'', Astropart. Phys. {\bf 25}, 179 (2006)
\bibitem{Nantais} C. Nantais, ``Radiopurity measurement of acrylic for the DEAP-3600 dark matter experiment'', M.Sc. thesis, Queen's University (2014)

\end{thebibliography}
\end{document}